\documentclass[12pt,preprint]{aastex}

\newcommand{\etal}{et~al.\ }

\newcommand{\pflux}{\hbox{photon~cm$^{-2}$~s$^{-1}$}}

\newcommand{\lumin}{\hbox{erg~s$^{-1}$}}

\newcommand{\be}{\begin{equation}}
\newcommand{\ee}{\end{equation}}
\newcommand{\ba}{\begin{eqnarray}}
\newcommand{\ea}{\end{eqnarray}}

\newcommand{\batse}{BATSE}
\newcommand{\swift}{\emph{Swift}}

\begin{document}

\def\Mesz{M\'esz\'aros}
\def\sarc{$^{\prime\prime}\!\!.$}
\def\arcsec{$^{\prime\prime}$}
\def\ls{\lower 2pt \hbox{$\;\scriptscriptstyle \buildrel<\over\sim\;$}}
\def\gs{\lower 2pt \hbox{$\;\scriptscriptstyle \buildrel>\over\sim\;$}}

\title{Intensity Distribution and Luminosity Function of the \emph{Swift} Gamma-Ray Bursts}

\author{Xinyu Dai\altaffilmark{1}} 

\altaffiltext{1}{Department of Astronomy,
University of Michigan, Ann Arbor, MI 48109,
xdai@umich.edu}

\begin{abstract}
	Using the sample of long Gamma-ray bursts (GRBs) detected by \swift-BAT before June 2007, we measure the cumulative
	distribution of the peak photon fluxes ($\log N$--$\log P$) of the \swift\ bursts.
	Compared with the \batse\ sample, we find that the two distributions are consistent after correcting the 
	band pass difference, suggesting that the two instruments are sampling the same population of bursts. 
	We also compare the $\log N$--$\log P$ distributions for sub-samples of the \swift\ bursts, 
	and find evidence for a deficit (99.75\% confident) of dark bursts without optical counterparts 
	at high peak flux levels,
	suggesting different redshift or $\gamma$-ray luminosity distributions for these bursts.
	The consistency between the $\log N$--$\log P$ distributions for the optically detected bursts with and without redshift measurements indicates that the current sample of the \swift\ bursts with redshift measurements, although selected heterogeneously, represents a fair sample of the non-dark bursts.  
	We calculate the luminosity functions of this sample in two redshift bins ($z<1$ and $z\ge1$), 
	and find a broken power-law is needed to fit the low redshift bin,
	where $dN/dL \propto L^{-1.27\pm0.06}$ for high luminosities ($L_{peak} > 5\times10^{48}\lumin$) 
	and $dN/dL \propto L^{-2.3\pm0.3}$ at for low luminosities, confirming the results of several studies for 
	a population of low luminosity GRBs.  
\end{abstract}


\section{Introduction}
The cumulative distribution of source intensities ($\log N$--$\log S$) is a useful tool for studying the source populations, especially when the redshifts of the sources are not measured.
In the Gamma-ray burst (GRB) field, since the peak photon flux of the GRB is directly related to the detection threshold, the
cumulative distribution of peak photon flux ($\log N$--$\log P$) is used in many studies.
A number of these studies have made use of the \batse\ sample of more than 2,000 GRBs.
This sample was first used to demonstrate the cosmological nature of the bursts (e.g., Fenimore et al.\ 1993; Pendleton et al.\ 1996), 
and then used to constrain the GRB populations (e.g., Kommers et al.\ 2000; Lin et al.\ 2004; Guetta et al.\ 2005; Dai \& Zhang 2005), though many assumed that GRB rate follows the star formation rate in the analyses.
Since the launch of \swift\ (Gehrels et al.\ 2004), the sample of the \swift-BAT bursts has reached to 237 in June 2007 (Sakamoto et al.\ 2008).
With the sample size, it is now possible to compare the \swift\ and \batse\ samples to test whether the two instruments sample the same population of bursts. 
It is possible that they are different since the two instruments have different band passes and sensitivities. 
For example, Band (2006) showed that \swift\ is more sensitive to soft bursts.
Moreover, there is still a large number of bursts without redshift measurements, e.g., dark bursts, and the $\log N$--$\log P$ distribution
provides a means of studying their source population besides using pseudo-redshifts derived from spectral or timing properties (e.g., Norris 2002).

The fraction of the \swift\ bursts with redshift measurements has increased significantly compared to the \batse\ bursts.
Although selected heterogeneously, it is tempting to measure the luminosity function of the \swift\ bursts using this
sample (e.g., Liang et al.\ 2007).  Besides issues with redshift selection effects, 
the \swift\ trigger efficiency has not been well studied, which presents
an additional difficulty.  
We show that by studying the $\log N$--$\log P$ distributions of the \swift\ and \batse\ bursts, 
we can justify the usage of the heterogeneous redshift sample and set detection thresholds for measuring
luminosity functions.  We present the luminosity functions using the heterogeneous redshift sample, where
we adopt a cosmology of $H_0 = 70~\rm{km~s^{-1}~Mpc^{-1}}$, $\Omega_{\rm m} = 0.3$, 
and $\Omega_{\Lambda}= 0.7$.

\section{The \swift\ Burst Sample}
We use the \swift-BAT GRB catalog published in Sakomoto et al. (2008).  The sample consists of 237 \swift\
bursts detected before 2007 June 16.  All the bursts are triggered by the BAT instrument on board \swift. 
The \swift-BAT catalog contains a number of basic properties of the bursts such as the burst duration, spectral
index, and peak photon flux in several bands.  There are 229 bursts with peak photon flux estimates in the 15--150 keV band,
with a minimum value of 0.23~\pflux.  Of these 229 bursts, 210 bursts can be identified as
long bursts and 15 as short bursts, where we divide the sample at $T_{90} = 2$~s.
In this paper, we focus on these 210 long bursts.  
To compare with the \batse\ burst sample, we use the long \batse\ burst sample from Kommers et al. (2000), which consists of   
2176 long GRBs from both online and off-line searches.

\begin{figure}
	\epsscale{1}
	\plotone{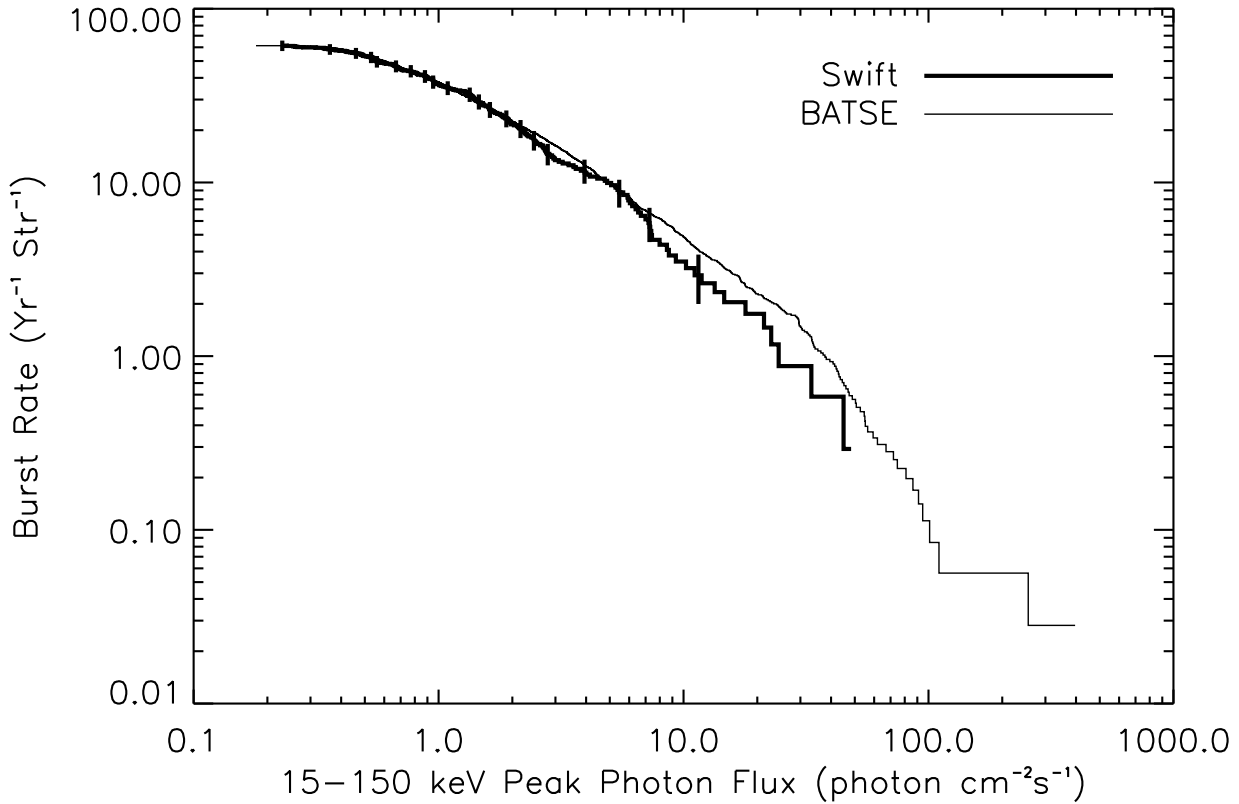}
	\caption{The $\log N$--$\log P$ distributions of the \swift\ bursts (thick solid line) and \batse\ bursts (thin solid line).
	We use the \swift\ sample from Sakamoto et al.\ (2008) and the \batse\ sample from Kommers et al.\ (2000).
	The peak flux values of the \batse\ bursts are corrected to the BAT band pass of 15--150 keV using an average correction factor of $N_{50-300} = 0.4*N_{15-150}$.
	The K-S test shows that the null probability that the two samples are drawn from the same distribution is 69\%, which indicates that the BAT and \batse\ instruments sample the same population of GRBs.
	\label{fig:np}}
\end{figure}

\section{The $\log N$--$\log P$ Distribution}
We plot the $\log N$--$\log P$ distributions of the \swift\ and \batse\ bursts in Figure~\ref{fig:np}.
We use the $>50$\% coded field of view of $\Omega_{\swift}=1.4$~str for the BAT and a flight time of $T_{\swift}=2.49$~yr until 2007 June 16 to calculate the burst rate for the \swift\ sample.
For the \batse\ sample, we adopt the values from Kommers et al.\ (2000) with $T_{\batse} = 1.33\times10^{8}$~s
and a mean field of view of $\Omega_{\batse} = 0.67*4\pi$~str. 
We also correct for the band pass differences between the BAT (15--150 keV) and \batse\ (50--300 keV) instruments,
where we use the spectral fits provided by Sakamoto et al.\ (2008) for the \swift\ sample.  
Using the simple power-law fits to the BAT spectra, which fit well for most \swift\ bursts (Sakamoto et al.\ 2008), 
we find a mean relation, $N_{50-300} = 0.4*N_{15-150}$, which we use to correct the photon
flux for the \batse\ bursts. 
Figure~\ref{fig:np} shows that the two distributions are quite consistent, and only small discrepancies with 
$\sim1\sigma$ significance exist at high photon flux levels at $f_{peak} > 8~\pflux$.  
We perform a Kolmogorov-Smirnov (K-S) test for the two samples,
and find that the null probability that the two samples are drawn from the same distribution is 69\% (76\% for 
the sub-samples with $f_{peak} > 8~\pflux$). 
This result indicates that the BAT and \batse\ instruments sample the same population of GRBs, even though they have different band passes.
We also convert the BAT photon flux to the \batse\ band individually for each burst using the simple power-law fits
of Sakamoto et al.\ (2008) to reduce the uncertainty introduced by scatter of the spectral indices around their mean, 
and compare the two distributions in the \batse\ band.  
We again find that the two distributions are not significantly different.
For some bursts, the spectra are better fit by a cut-off power-law (Sakamoto et al.\ 2008), which introduces 
additional uncertainties in the flux conversion; however they are generally negligible compared to the uncertainty introduced by scatter of the photon indices. 
We note that the detection limits of the two samples also match each other after the band pass correction.

\begin{figure}
	\epsscale{1}
	\plotone{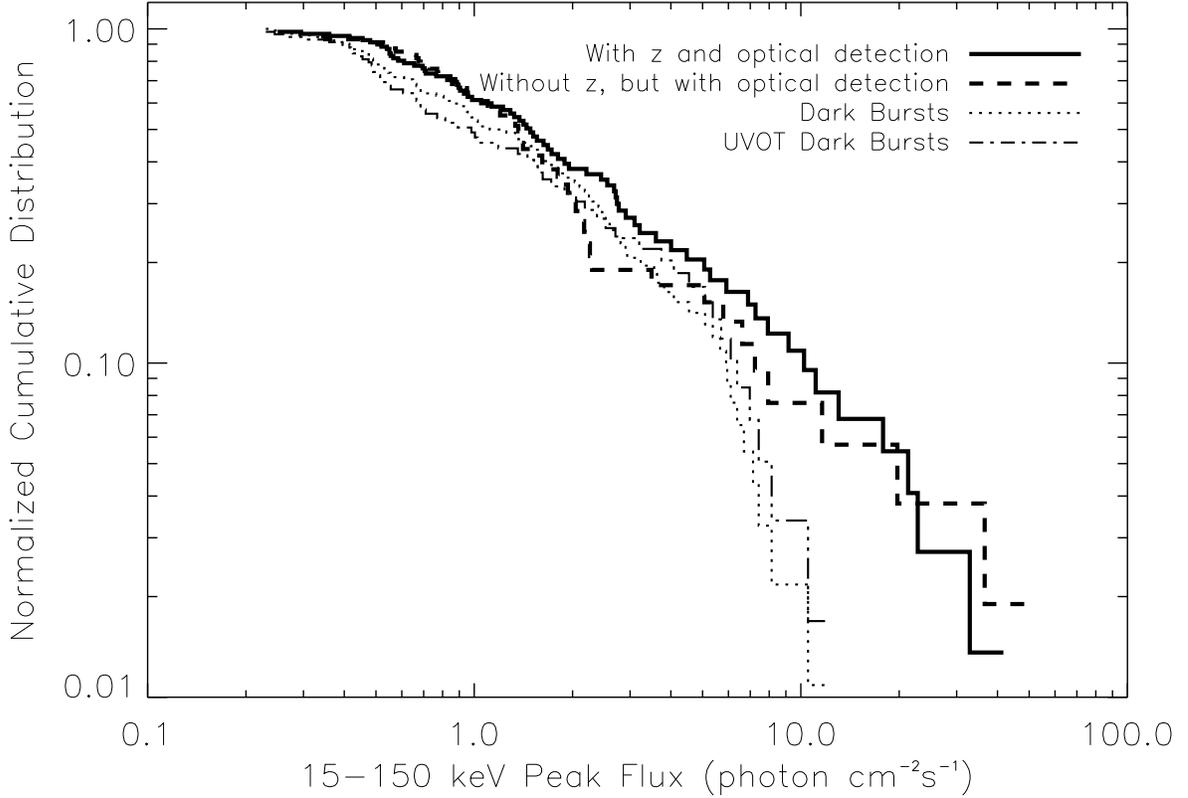}
	\caption{Comparisons between the $\log N$--$\log P$ distributions for the \swift\ sub-samples.
	The thick solid line is for the \swift\ bursts with optical detections and redshift measurements, and the thick dashed line is for bursts with optical detections but without redshift measurements.  The thin dotted line is for the dark bursts without
        optical detections, and the thin dash-dotted line is for bursts with UVOT observations and with Galactic extinction $A_{V, Gal}<1.5$~mag, but without optical detections.
	Although the K-S test results show that all four sub-samples are consistent with K-S probabilities of $\sim35$\%,
	there is an indication that the dark bursts are under-populated at high peak flux levels.
	The K-S test probability drops to 4\% when we compare the high peak flux bursts ($f_{peak} > 5~\pflux$) between the distributions for dark bursts and optically detected bursts.  
	At even higher peak flux levels ($f_{peak} > 10~\pflux$), the difference is more pronounced with a Poisson null
	probability of 0.25\%.
	\label{fig:snp}}
\end{figure}

\subsection{Comparisons between \swift\ Sub-Samples}
Next, we compare the $\log N$--$\log P$ distributions for sub-samples of the \swift\ bursts.  First, we compare
the first half and the second half of the \swift\ bursts separated by time, finding that the two samples are consistent, 
suggesting that there is no significant degradation in the BAT detector.  
Next, we compare the distributions for bursts with or without redshift measurements or optical detections (Figure~\ref{fig:snp}).
There are 68 bursts with redshift measurements and optical afterglow detections, 48 with optical afterglow 
detections but without redshift measurements, and 94 without optical afterglow detections.
The last category of bursts with no optical afterglow detections is the dark bursts, though another 
definition uses an optical to X-ray threshold ratio to define this sample (Jakobsson et al.\ 2004).
In this paper, we use the simple definition of no optical detection for dark bursts.
In most cases, we find no significant difference between the sub-samples based on the K-S test probabilities (45\% 
between samples with and without redshift measurements, 34\% between samples with redshift measurements and dark bursts,
and 24\% between samples with redshift measurements and without redshift measurements but with optical detections).

However, Figure~\ref{fig:snp} shows that there is an indication for a deficit of dark bursts at high peak flux levels compared to the bursts with optical afterglow detections.
We test it by including a flux filter and select bursts with $f_{peak} > 5~\pflux$, and find that the 
K-S probability that the dark bursts and bursts with optical detections are drawn from the same population is 4\%.
It seems that the deficit is more pronounced in the flux level at $f_{peak} > 10~\pflux$.  
Since the number of bursts detected in this regime is too small and that the K-S test is no longer applicable,  
 we use a simple Poisson argument instead.  
 The null model is that the dark bursts follow the same distribution as the optically
 detected bursts, which predicts that the number of the dark bursts detected at $f_{peak} > 10~\pflux$ is 8.2$\pm$0.8.
 This model prediction and the associated error-bar are obtained by aligning the normalizations of the two distributions at low flux levels.
Since we only detect one dark burst with $f_{peak} > 10~\pflux$, the Poisson probability of detecting no more than
one burst is 0.0025 for an expected value of 8.2.  If we use the lower end of the model prediction (7.4), the 
corresponding probability is 0.0051.
We argue that there is evidence that the dark bursts do not follow the $\log N$--$\log P$ distribution 
of the optically detected bursts.
It can be either a deficit of dark bursts at high peak flux levels or an over abundance of dark bursts at low
flux levels.
Future analysis including the whole \swift\ sample is needed to confirm this results.

Since the ground-based optical follow-up observations of GRBs can be potentially biased by several reasons, such as the scheduling issues
or unfavorable burst sky positions, we also test by limiting the dark burst sample using those bursts observed by UVOT but without UVOT or other optical detections (UVOT dark bursts).
This provides a more uniform selection because UVOT routinely observes the GRB fields after the burst triggers.
In addition, this excludes some of the bursts with unfavorable sky positions to follow-up in the optical bands, such as those located close to the Sun.
We also limit the sample by excluding bursts with large Galactic extinctions, $A_{V, Gal} > 1.5$~mag, where the dark burst fraction is significantly larger compared to that for the total population.
However, since the UVOT flux limit is generally shallower than those of
the ground-based observations using mid-sized telescopes, we still classify the dark bursts as those without optical detections.  We find the $\log N$--$\log P$ distribution of the UVOT dark 
burst sample follows that of the general dark burst sample (or no optical counterparts sample, Figure~\ref{fig:snp}).
This is because only a small fraction of bursts have unfavorable burst positions that cannot be observed by the
ground-based telescopes.
In addition, a number of optical telescopes are used to observe GRB afterglows world wide, which collectively
reduces the scheduling problems such as the whether conditions in individual optical sites.
Because of the shallowness of the UVOT flux limit, it is still possible that both of the dark burst (no optical counterpart) samples contain a fraction of normal, non-dark bursts, because they are not observed promptly by the mid or large size ground-based telescopes.  However, this population should not contribute
to the difference between the $\log N$--$\log P$ distributions between dark and normal bursts.

\section{Luminosity Function}
The bursts with redshift measurements account for 59\% of the bursts with optical afterglow detections.
Figure~\ref{fig:snp} shows that the $\log N$--$\log P$ distributions between the optically detected bursts with
and without redshift measurements are quite similar.  
This suggests that the current redshift sample, although obtained heterogeneously, represents a fair sample of the
bursts with optical afterglow detections.
We compute the luminosity function (LF) of this sample using the $1/V_{max}$ method (Schmidt 1968; Avni \& Bahcall 1980), where we include an additional factor of $1/(1+z)$ in the differential volume element to account
for the cosmological time dilation effect.
We calculate the k-corrections using the power-law fit to the BAT spectra from Sakomoto et al.\ (2008), and
use a detection limit of $f_{peak} = 0.25~\pflux$, the smallest value in the redshift sample (GRB~060218), 
to calculate the $V_{max}$ values. 
Since the $\log N$--$\log P$ distribution of the \swift\ sample matches that of the \batse\ sample 
when approaching the detection limits (Figure~\ref{fig:np}),
we use the trigger efficiency analysis of Kommers et al. (2000) to estimate the efficiency of the \swift\ triggers,
and then assign weights to the \swift\ bursts with a maximum weight of 2.
Using a maximum weight is a conservative approach to avoid huge corrections at low
peak flux levels, 
where the uncertainties of the efficiency analysis could be large.

\begin{figure}
	\epsscale{1}
	\plotone{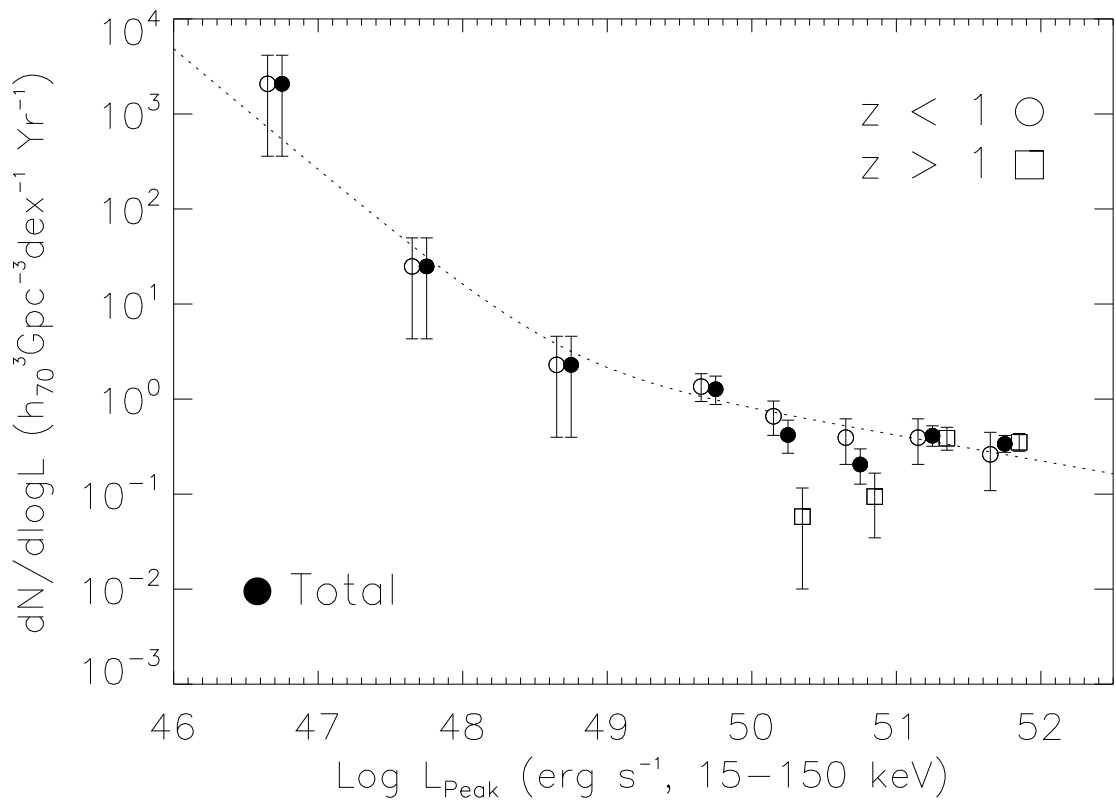}
	\caption{The luminosity function of the \swift\ bursts with redshift measurements.
	The data for the $z<1$ and $z\ge1$ bins are slightly shifted for clarity.
	The dashed dotted line is a broken power-law fit to the $z<1$ LF.
	\label{fig:lf}}
\end{figure}

We show the total LF and those in redshift bins of $z<1$ and $z\ge1$
in Figure~\ref{fig:lf}, where we choose a bin size of 0.5~dex.
We find that the LF for the lower redshift bin is consistent with the total LF,
and they cannot be fit well by a single power-law model with $\chi^2/dof =$ 4.6 and 2.8, respectively,
for the total and $z<1$ LFs.
We add another power-law component at the low luminosity end, and fit the $z<1$ LF.  
We find $dN/dL \propto L^{-2.3\pm0.3}$ at the low luminosity end and $dN/dL \propto L^{-1.27\pm0.06}$ at the high luminosity end ($L_{peak} > 5\times10^{48}\lumin$) with $\chi^2/dof =$1.3.
The complete LF for the $z<1$ bin is in Equation~\ref{eqn:lf},
\begin{equation}
	\frac{dN}{dL} = (1.7\pm0.2) {\rm h_{70}^3 Gpc^{-3} Yr^{-1}} \left[ \left(\frac{L}{5\times10^{48}{\rm erg~s^{-1}}}\right)^{-2.3\pm0.3}+\left(\frac{L}{5\times10^{48}{\rm erg~s^{-1}}}\right)^{-1.27\pm0.06} \right]\label{eqn:lf}
\end{equation}
Comparing the LFs in the two redshift bins, the two low luminosity data points for the $z\ge1$ LF 
at $L_{peak} = 10^{50-51}\lumin$ are below the $z<1$ LF.
Since the star formation rate drops significantly below $z=1$ (e.g., Hopkins \& Beacom 2006), if the GRB rate follows the star
formation rate, we expect that the LF for the $z\ge1$ bin should have a higher normalization than that for the $z<1$ LF.
It is possible that the two low luminosity data points of the $z\ge1$ LF are incomplete because they are close
to the \swift\ flux limit (e.g., Kistler et al.\ 2008),  
even though we have partially modeled the \swift\ trigger incompleteness by using the \batse\ trigger efficiency model (Kommers et al.\ 2000).
If we neglect these two low luminosity data points, 
the remaining two data points of the $z\ge1$ LF are not lower than the $z<1$ LF. 
Considering the uncertainties, the results do not present a significant challenge to the hypothesis that 
the GRB rate follows the star formation rate.
We note that the LFs do not include the contribution from the dark bursts.  We need to 
multiply by a factor of 1.8 to include them; however, 
this may not be accurate since the dark bursts may
have a different redshift or $\gamma$-ray luminosity distribution (\S3.1).
In addition, if the majority of the dark bursts are at high redshifts (\S5), the normalization of the $z\ge1$ 
LF will be significantly increased compared to the $z<1$ LF, which can raise the GRB rate to be consistent
with the star formation rate, or even evolving faster as suggested by some recent studies (e.g., Salvaterra \& Chinarini 2007;
Kistler et al.\ 2008; Salvaterra et al.\ 2009).

\section{Discussion}
We compare the $\log N$--$\log P$ distributions for the \swift\ bursts and \batse\ bursts, and find that they
are consistent after correcting the band pass differences as suggested by the K-S test results.  
This shows that the two instruments sample
the same population of bursts, although they have different band passes.
We can also compare the normalizations of the distributions directly, since 
the field of view and the observing time of the two samples are measured,  
and we find that they are consistent.
Indirect comparison between the \swift\ and \batse\ samples has been performed in previous studies (Salvaterra \& Chincarini 2007; Virgili et al.\ 2009), where the two samples are found to be consistent with same theoretical models.
Given that \swift\ is more sensitive than \batse\ for the soft bursts (Band 2006), it is puzzling that 
the two distributions are consistent.  Currently, we are comparing the on-board triggered \swift\ sample with the \batse\ off-line search sample. It is possible that the difference will become significant when comparing both
the off-line search samples.

We also compare the $\log N$--$\log P$ distributions for the sub-samples of the \swift\ bursts with or without
optical afterglow detections or redshift measurements.  We find them to be broadly consistent.
However, at the high peak flux regime ($f_{peak} > 5~\pflux$), we find an indication that there is a deficit 
(96\% confident) of ``dark bursts'' (bursts without optical counterparts).
This deficit is more evident at $f_{peak} > 10~\pflux$, where we find the effect is significant at the 99.75\%
confidence level.

A comparison between the $\log N$--$\log P$ distributions for dark bursts and optically bright bursts can 
place important constraints on the origin of the 
dark bursts.  There are several models proposed for the dark bursts.  
They can be bursts with intrinsic normal optical-to-$\gamma$-ray ratios but with large intrinsic or foreground optical extinction (e.g., Taylor et al.\ 1998; Djorgovski et al.\ 2001; Fynbo et al.\ 2001),
 have intrinsic faint optical afterglows, i.e., low optical-to-$\gamma$-ray ratios, (e.g., Groot et al.\ 1998a; Frail et al.\ 1999), are normal bursts but exist in extremely high redshifts where the Lyman break lands in the optical (e.g., Groot et al.\ 1998b; Fynbo et al.\ 2001),
 or a combination of several origins mentioned above.
A different $\gamma$-ray luminosity distribution or redshift distribution is needed to interpret the difference in the 
$\log N$--$\log P$ distributions.
Since the $\gamma$-ray/hard X-ray flux is not sensitive to the ISM absorption, the hypotheses with optical
extinction or intrinsic optical faintness predict that the $\log N$--$\log P$ distributions should have a similar shape
between dark and optically detected bursts.  For the high redshift scenario, it is possible for the dark bursts to have a different
shape in the $\log N$--$\log P$ distribution from the normal, non-dark bursts.  Therefore, the deficit of the 
dark bursts at high flux levels suggests that a significant fraction
of dark burst is from extreme high redshifts.
In addition, it is possible that the bulk of the dark bursts are selected due to their low optical
luminosity, and the deficit of dark bursts at high peak photon fluxes merely reflects the detection
threshold of optical observations under the assumption of a constant optical-to-$\gamma$-ray ratio.

We measure the luminosity function of the \swift\ bursts using the sample with redshift measurements.  
We find the luminosity function can be fit by a broken power-law with $dN/dL \propto L^{-2.3\pm0.3}$ at the low
luminosity end, and $dN/dL \propto L^{-1.27\pm0.06}$ at the high luminosity end. 
We compare this result with previous measurements (Schmidt 2001; Norris 2002; Stern et al.\ 2002;
Firmani et al.\ 2004; Guetta et al.\ 2005; Liang et al.\ 2007; Salvaterra \& Chincarini 2007) and find a best match with the result
from Liang et al.\ (2007).
The requirement for an additional component at the low luminosity end confirms the 
existence of the population of low luminosity GRBs claimed by several studies (e.g., Cobb et al.\ 2006; 
Piran et al.\ 2006; Soderberg et al.\ 2006; Guetta \& Della Valle 2007; Liang et al.\ 2007; Virgili et al.\ 2009).
Although we model it as a power-law, it could be the tail of a Gaussian component.
At the high luminosity range, the measured slope $dN/dL \propto L^{-1.27\pm0.06}$ is close to the 
prediction of the 
``quasi-universal Gaussian jet'' ($dN/dL \propto L^{-1}$, Zhang et al.\ 2004).
At the very high luminosity end ($L_{peak} > 10^{51}\lumin$), the ``quasi-universal Gaussian jet'' predicts
$dN/dL \propto L^{-2}$ (Lloyd-Ronning et al.\ 2004; Dai \& Zhang 2004), which cannot be tested with the current data.

Unlike many of the previous studies, we do not make any assumption on the GRB rate, since there are recent
studies suggesting that the GRB rate does not follow the star formation rate (e.g., Stanek et al.\ 2006).
Instead, we measure the average GRB luminosity function in a large redshift bin.  
We compare the LFs in two luminosity bins.  The $z\ge1$ LF shows a drop at two low luminosity data points.  
This result can be affected by the uncertainties in the \swift\ trigger efficiency close to the \swift\
detection limit.
If we neglect these two data points, the LFs for the two redshift bins are consistent.
However, the large measurement uncertainties in the $z<1$ LF make it difficult to test whether the
GRB rate follows the star formation rate.

\acknowledgements 
We thank B. Zhang, C. S. Kochanek, R. Salvaterra, and the anonymous referee for helpful discussion.

\clearpage

\end{document}